\documentclass[12pt,preprint]{aastex}
 \usepackage{lineno}
 \usepackage{multirow}
 \usepackage{color}
 \usepackage{rotating}

\shorttitle{1A~0535+262 during a giant X-ray outburst}
\shortauthors{VERITAS et al.}

\begin{document}
\title{Gamma-ray observations of the Be/pulsar binary 1A~0535+262 during a giant X-ray outburst}

\author{
V.~A.~Acciari\altaffilmark{1},
E.~Aliu\altaffilmark{2},
M.~Araya\altaffilmark{3},
T.~Arlen\altaffilmark{4},
T.~Aune\altaffilmark{5},
M.~Beilicke\altaffilmark{6},
W.~Benbow\altaffilmark{1},
S.~M.~Bradbury\altaffilmark{7},
J.~H.~Buckley\altaffilmark{6},
V.~Bugaev\altaffilmark{6},
K.~Byrum\altaffilmark{8},
A.~Cannon\altaffilmark{9},
A.~Cesarini\altaffilmark{10},
L.~Ciupik\altaffilmark{11},
E.~Collins-Hughes\altaffilmark{9},
W.~Cui\altaffilmark{3,\dag},
R.~Dickherber\altaffilmark{6},
C.~Duke\altaffilmark{12},
A.~Falcone\altaffilmark{13},
J.~P.~Finley\altaffilmark{3},
L.~Fortson\altaffilmark{11},
A.~Furniss\altaffilmark{5},
N.~Galante\altaffilmark{1},
D.~Gall\altaffilmark{3},
S.~Godambe\altaffilmark{14},
S.~Griffin\altaffilmark{15},
R.~Guenette\altaffilmark{15},
G.~Gyuk\altaffilmark{11},
D.~Hanna\altaffilmark{15},
J.~Holder\altaffilmark{16},
G.~Hughes\altaffilmark{21},
C.~M.~Hui\altaffilmark{14},
T.~B.~Humensky\altaffilmark{17},
A.~Imran\altaffilmark{18},
P.~Kaaret\altaffilmark{19},
M.~Kertzman\altaffilmark{20},
H.~Krawczynski\altaffilmark{6},
F.~Krennrich\altaffilmark{18},
A.~S.~Madhavan\altaffilmark{18},
G.~Maier\altaffilmark{21,\dag},
P.~Majumdar\altaffilmark{4},
S.~McArthur\altaffilmark{6},
P.~Moriarty\altaffilmark{22},
R.~A.~Ong\altaffilmark{4},
A.~N.~Otte\altaffilmark{5},
D.~Pandel\altaffilmark{19},
N.~Park\altaffilmark{17},
J.~S.~Perkins\altaffilmark{1},
M.~Pohl\altaffilmark{23},
H.~Prokoph\altaffilmark{21},
J.~Quinn\altaffilmark{9},
K.~Ragan\altaffilmark{15},
L.~C.~Reyes\altaffilmark{17},
P.~T.~Reynolds\altaffilmark{24},
E.~Roache\altaffilmark{1},
H.~J.~Rose\altaffilmark{7},
D.~B.~Saxon\altaffilmark{16},
G.~H.~Sembroski\altaffilmark{3},
G.~Demet~Senturk\altaffilmark{25},
A.~W.~Smith\altaffilmark{8},
G.~Te\v{s}i\'{c}\altaffilmark{15},
M.~Theiling\altaffilmark{1},
S.~Thibadeau\altaffilmark{6},
A.~Varlotta\altaffilmark{3,\dag},
S.~Vincent\altaffilmark{14},
M.~Vivier\altaffilmark{16},
S.~P.~Wakely\altaffilmark{17},
J.~E.~Ward\altaffilmark{9},
T.~C.~Weekes\altaffilmark{1},
A.~Weinstein\altaffilmark{4},
T.~Weisgarber\altaffilmark{17},
S.~Weng\altaffilmark{3},
D.~A.~Williams\altaffilmark{5},
M.~Wood\altaffilmark{4},
B.~Zitzer\altaffilmark{3}}

\altaffiltext{1}{Fred Lawrence Whipple Observatory, Harvard-Smithsonian Center for Astrophysics, Amado, AZ 85645, USA}
\altaffiltext{2}{Department of Physics and Astronomy, Barnard College, Columbia University, NY 10027, USA}
\altaffiltext{3}{Department of Physics, Purdue University, West Lafayette, IN 47907, USA; avarlott@purdue.edu, cui@purdue.edu}
\altaffiltext{4}{Department of Physics and Astronomy, University of California, Los Angeles, CA 90095, USA}
\altaffiltext{5}{Santa Cruz Institute for Particle Physics and Department of Physics, University of California, Santa Cruz, CA 95064, USA}
\altaffiltext{6}{Department of Physics, Washington University, St. Louis, MO 63130, USA}
\altaffiltext{7}{School of Physics and Astronomy, University of Leeds, Leeds, LS2 9JT, UK}
\altaffiltext{8}{Argonne National Laboratory, 9700 S. Cass Avenue, Argonne, IL 60439, USA}
\altaffiltext{9}{School of Physics, University College Dublin, Belfield, Dublin 4, Ireland}
\altaffiltext{10}{School of Physics, National University of Ireland Galway, University Road, Galway, Ireland}
\altaffiltext{11}{Astronomy Department, Adler Planetarium and Astronomy Museum, Chicago, IL 60605, USA}
\altaffiltext{12}{Department of Physics, Grinnell College, Grinnell, IA 50112-1690, USA}
\altaffiltext{13}{Department of Astronomy and Astrophysics, 525 Davey Lab, Pennsylvania State University, University Park, PA 16802, USA}
\altaffiltext{14}{Department of Physics and Astronomy, University of Utah, Salt Lake City, UT 84112, USA}
\altaffiltext{15}{Physics Department, McGill University, Montreal, QC H3A 2T8, Canada}
\altaffiltext{16}{Department of Physics and Astronomy and the Bartol Research Institute, University of Delaware, Newark, DE 19716, USA}
\altaffiltext{17}{Enrico Fermi Institute, University of Chicago, Chicago, IL 60637, USA}
\altaffiltext{18}{Department of Physics and Astronomy, Iowa State University, Ames, IA 50011, USA}
\altaffiltext{19}{Department of Physics and Astronomy, University of Iowa, Van Allen Hall, Iowa City, IA 52242, USA}
\altaffiltext{20}{Department of Physics and Astronomy, DePauw University, Greencastle, IN 46135-0037, USA}
\altaffiltext{21}{DESY, Platanenallee 6, 15738 Zeuthen, Germany; gernot.maier@desy.de}
\altaffiltext{22}{Department of Life and Physical Sciences, Galway-Mayo Institute of Technology, Dublin Road, Galway, Ireland}
\altaffiltext{23}{Institut f\"ur Physik und Astronomie, Universit\"at Potsdam, 14476 Potsdam-Golm,Germany}
\altaffiltext{24}{Department of Applied Physics and Instrumentation, Cork Institute of Technology, Bishopstown, Cork, Ireland}
\altaffiltext{25}{Columbia Astrophysics Laboratory, Columbia University, New York, NY 10027, USA}

\altaffiltext{\dag}{Authors to whom any correspondence should be addressed.}

\begin{abstract}

Giant X-ray outbursts, with luminosities of about $ 10^{37}$ erg s$^{-1}$, are observed roughly every 5 years from the nearby Be/pulsar binary 1A~0535+262. In this article, we present observations of the source with VERITAS at very-high energies (VHE; E$>$100 GeV) triggered by the X-ray outburst in December 2009. The observations started shortly after the onset of the outburst, and they provided comprehensive coverage of the episode, as well as the 111-day binary orbit. No VHE emission is evident at any time. 
We also examined data from the contemporaneous observations of 1A~0535+262 with the Fermi/LAT at high energy photons (HE; E$>$0.1 GeV) and failed to detect the source at GeV energies. The X-ray continua measured with the Swift/XRT and the RXTE/PCA can be well described by the combination of blackbody and Comptonized emission from thermal electrons. Therefore, the gamma-ray and X-ray observations suggest the absence of a significant population of non-thermal particles in the system. This distinguishes 1A~0535+262 from those Be X-ray binaries (such as PSR~B1259--63 and LS~I~+61$^{\circ}$303) that have been detected at GeV--TeV energies. We discuss the implications of the results on theoretical models. 

\end{abstract}

\keywords{acceleration of particles Ð binaries: general - gamma rays: observations - individual (1A~0535+262)}

\section{Introduction}

High-mass X-ray binaries are binary systems in which the mass of the secondary star exceeds 10 solar masses. The majority are Be/X-ray binaries (BeXRBs), consisting of a Be or O star and a neutron star \citep{Liu-2006}. Some of the BeXRBs manifest themselves observationally as X-ray pulsars. Be or Oe stars are known to be rapid rotators, so their winds are concentrated strongly around the equatorial plane, forming a circumstellar disk (see review by \citet{1995xrbi.nasa...58V}). The binary orbits of BeXRBs tend to be quite wide and eccentric, with orbital periods typically exceeding 20 days. 


%
BeXRBs are complex systems.  They exhibit flux variability over a wide range of wavelengths. The variability can be regular, often correlated with the orbital movement of the neutron star around the OBe star, or in sudden and strong outbursts. The orbital modulation may be caused by: the varying geometry of the binary system, changes in the physical conditions of the circumstellar disk of the Be star, or the occasional formation of an accretion disk around the compact object. Normal (or Type I) outbursts, with luminosities of $<10^{37}$ erg s$^{-1}$, are thought to occur when the neutron star in a BeXRB passes through the circumstellar disk (near periastron) and thus accretes at a higher rate \citep{Reig-2007}. Giant (or Type II) outbursts occur more rarely but may be seen at any orbital phase, with X-ray luminosities close to the Eddington limit; they are thought to be associated with the formation of a transient accretion disk \citep{Stella-1986}.
 
%


Recently, several high-mass X-ray binaries have been seen to emit photons in the very-high energy range. Among them, PSR~B1259--63/LS~2883 \citep{Aharonian-2005a}
 and LS~I~+61$^{\circ}$303 \citep{Albert-2006, Acciari-2008} are BeXRBs and are thought to have circumstellar disks. A third VHE source, HESS~J0632+057, may also fall in this VHE binary class \citep{Aharonian-2004, Hinton-2009, Acciari-2009}. The production of VHE gamma rays can be envisioned in both leptonic and hadronic scenarios. Leptonic models postulate that the interaction between pulsar wind and stellar wind may result in the formation of a strong shock that could accelerate electrons to relativistic energies and that the relativistic electrons may produce VHE gamma rays by upscattering ambient photons. In hadronic scenarios (e.g.~\cite{Romero-2001} and \cite{Orellana-2005}),
a hadron beam is accelerated in the magnetosphere of a neutron star. It impacts the transient accretion disk and produces VHE gamma rays via the decay of neutral pions produced in pp-interactions, a mechanism first proposed by \citet{Cheng-1989, Cheng-1991}. 

VHE gamma rays are expected to interact with ambient photons to produce e$^{+}$e$^{-}$ pairs, leading to orbital modulation of the VHE gamma-ray flux \citep{Bednarek-1993}. During an outburst, the VHE gamma-ray flux would then peak at the beginning or end of the outburst, when thermal emission from the accretion disk is at a minimum;  the system is opaque to gamma rays near the peak of the outburst \citep{Romero-2001,Orellana-2007}. In hadronic scenarios, the gamma-rays emitted by the secondary electrons and positrons that are produced in electromagnetic cascades may contribute to fluxes in the MeV to GeV range.

1A~0535+262 has been extensively observed over the years, due to the many giant and minor outbursts that have occurred. The pulsar shows evidence for three representative X-ray intensity states, one of quiescence with a flux level of $\lesssim$10 mCrab, a normal outburst with a flux level of  $\sim$10 mCrab to 1 Crab, and a giant outburst with a flux level of several Crab \citep{Kendziorra-1994}. Giant outbursts have been observed in April 1975 \citep{Rosenberg-1975}, October 1980 \citep{Estulin-1981}, June 1983 \citep{Sembay-1990}, March/April 1989 \citep{Makino-1989}, February 1994 \citep{Finger-1996}, May/June 2005 \citep{Caballero-2007a} and December 2009 \citep{Wilson-Hodge-2010}. Studies have been conducted on recent minor outbursts as well \citep{Naik-2005,Caballero-2010}. 
An extensive review of the system was provided by \cite{Giovanelli-1992}.

\begin{figure}
\plotone{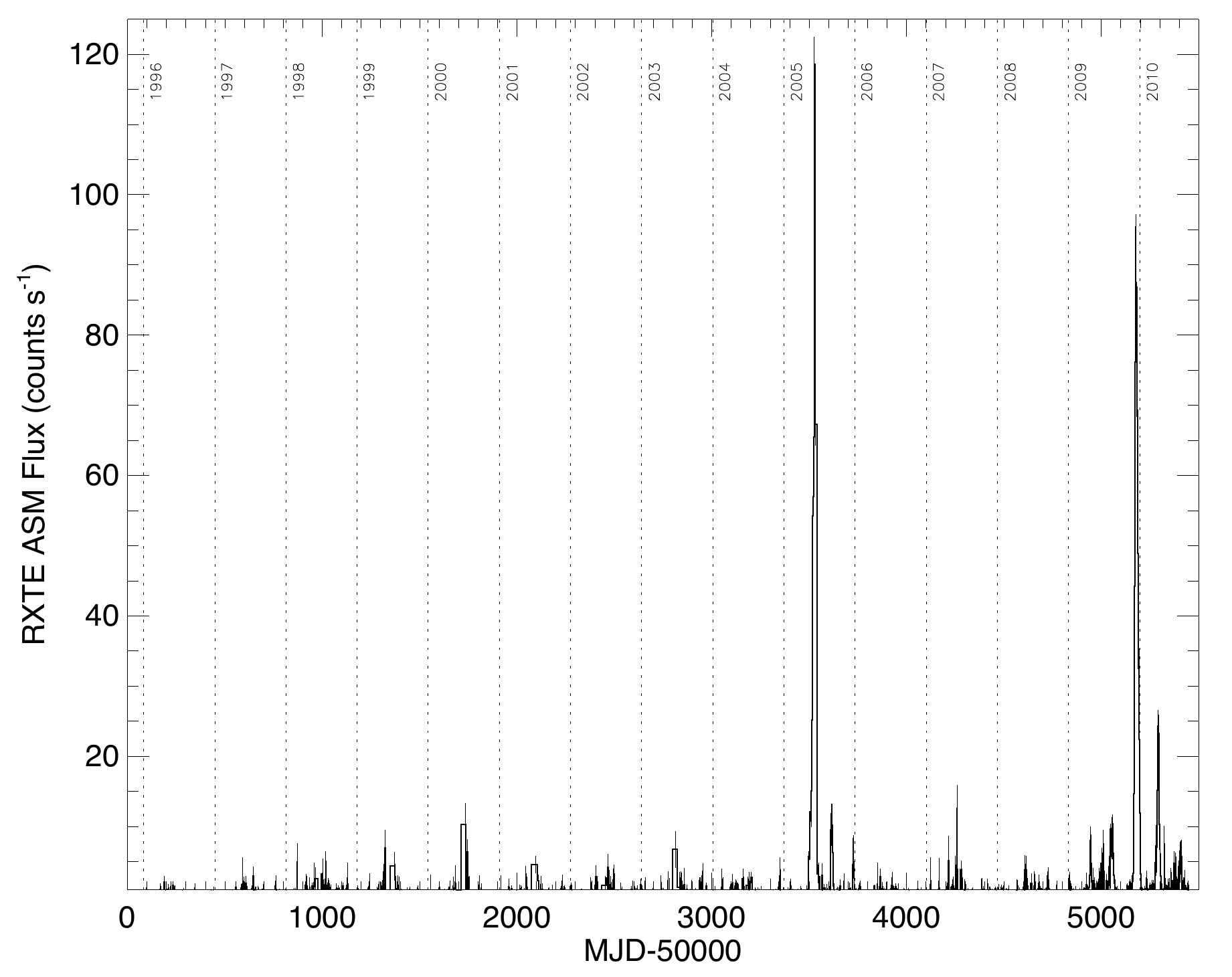}
\caption{\label{fig:asmlc}
Long-term RXTE/ASM (1.5--12 keV) light curve of 1A~0535+262, from 1996 January 5 to 2010 September 9. Note the presence of outbursts of different types.
}
\end{figure}

The Be/X-ray binary 1A~0535+262 has been associated with the EGRET unidentified gamma-ray source 3EG J0542+2610 \citep{Romero-2001} and has therefore long been considered as an object of interest for VHE gamma-ray observations. 
At a distance of about 2 kpc, 1A~0535+262 is among the closest binary systems \citep{Steele-1998}. It consists of a pulsar of period 103 s in an eccentric orbit ($e=0.47$) with an O9.7-B0 IIIe star. The orbital period is 111 days \citep{Coe-2006}. X-ray observations of 1A~0535+262 may suggest 
the formation of a transient accretion disk during giant outbursts \citep{Finger-1996}. Figure~\ref{fig:asmlc} shows a long-term X-ray light curve of 1A~0535+262, as measured with the All-Sky Monitor (ASM) on board the {\it Rossi X-ray Timing Explorer} (RXTE), over a period of roughly 15 years. The presence of both normal and giant outbursts is clearly seen. The December 2009 outburst peaked around 1.3 Crab (95 ASM counts s$^{-1}$) and it is second only to the May/June 2005 outburst peak (1.7 Crab, 120 counts s$^{-1}$). The outburst duration, $\sim$40 days, the X-ray flux intensity (peak intensity greater than the Crab) and the occurrence of the outburst at a random binary phase are all in line with a representative Type II outburst event.
%

\begin{figure}
\plottwo{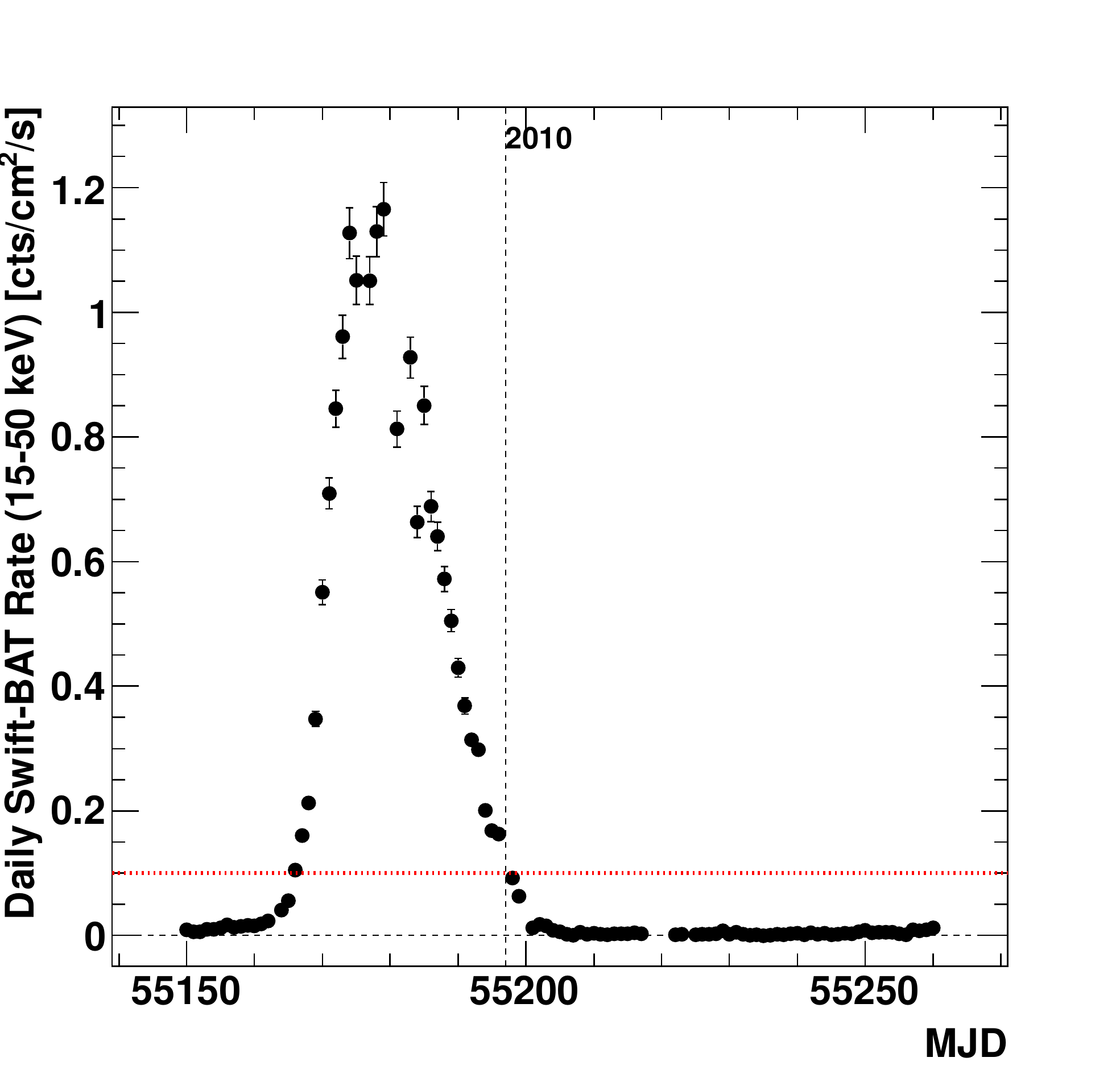}{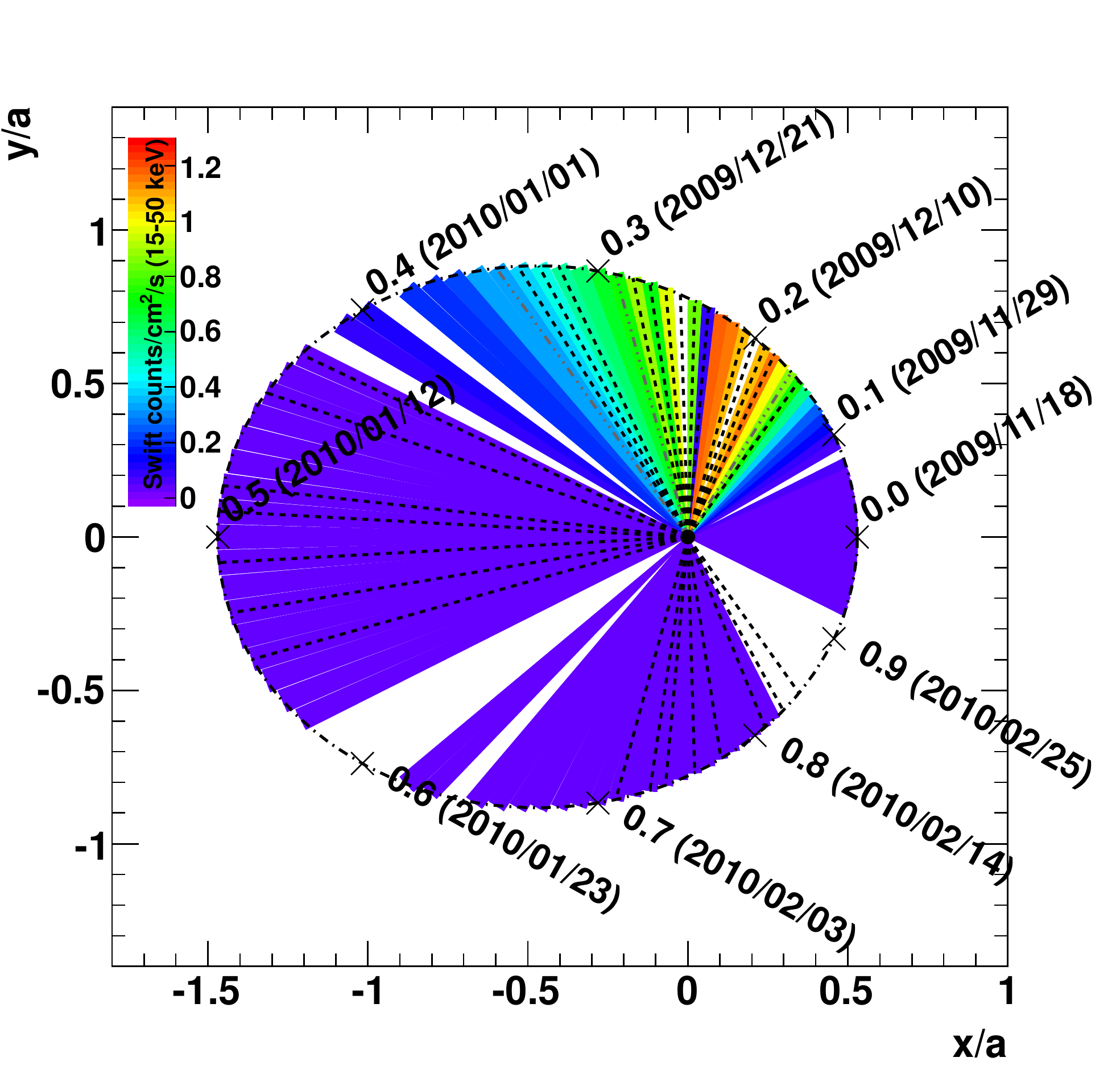}                                                                      
\caption{\label{fig:Slightcurve}
Left: Swift/BAT counting rate vs time in the 15-50 keV energy range.
The horizontal line at 0.1 cts/cm$^2$/s indicates the trigger threshold for observations with VERITAS.
Right: Relative orbit of the neutron star around the Be star.
The primary star lies in the focus of the ellipse (0,0) and the axis units are multiples of the semi-major axis of the orbit. 
The most probably inclination of the binary system is 35 to $39^{\circ}$ \citep{Giovanelli-2007}.
Indicated in colors is the Swift/BAT counting
rate in the 15-50 keV energy range
for the orbit starting in November 2009.
The dashed lines indicate nights with VERITAS observations, covering the flare, apastron, and periods close to periastron.
Orbital parameters after \cite{Coe-2006}.
}
\end{figure}

In this article we report on X-ray and gamma-ray observations of 1A~0535+262 during a giant outburst in December 2009 (as shown in Fig.~\ref{fig:asmlc}). During the outburst, the hard X-ray (15-50 keV) flux, as measured with the Burst Alert Telescope (BAT) on board the {\it Swift Gamma Ray Burst Mission} (Swift), reached a level of $>$5 times that of the Crab Nebula. Triggered by the alerts from the Swift/BAT, the VHE gamma-ray observatory, the {\it Very Energetic Radiation Imaging Telescope Array System} (VERITAS), monitored 1A~0535+262 for a full orbital period (see Fig.~\ref{fig:Slightcurve}). These data are complemented by observations in soft X rays with the X-ray Telescope (XRT) on Swift, in hard X rays with the Proportional Counter Array (PCA) on RXTE, and in GeV gamma rays with the Large Area Telescope (LAT) on {\it Fermi Gamma Ray Space Telescope}. 

\section{Observations}

\subsection{VERITAS: VHE Gamma-Ray Observations}

VERITAS is a ground-based gamma-ray telescope array located at the Fred Lawrence Whipple Observatory in southern Arizona (1.3 km above sea level,	N31$^{\mathrm{o}}$ 40', W110$^{\mathrm{o}}$ 57'). It consists of four 12-meter imaging Cherenkov telescopes with 499-pixel cameras, designed to detect the faint flashes of Cherenkov light from air showers initiated in the atmosphere by high-energy photons. VERITAS is sensitive to photons in the energy range from 100 GeV to 30 TeV, with a maximum effective area of approximately 10$^5$ m$^2$. At 1 TeV the angular resolution is better than 0.1$^{\mathrm{o}}$ and the energy resolution is 
15-20\%. The field of view of the VERITAS telescopes is 3.5$^{\mathrm{o}}$. The observatory was upgraded in summer 2009, leading to a sensitivity sufficient to detect  sources with a flux of 1\% of the Crab Nebula in less than 30 hours of observations (1 h for 5\% of the Crab Nebula flux). Observations with VERITAS are possible during dark-sky and moderate moonlight conditions (the moon less than 50\% illuminated). Elevated background light levels during moderate moonlight lead to a lower sensitivity to gamma rays at the threshold. For more details on the VERITAS instrument, see e.g. \cite{Acciari-2008}.

VERITAS observed 1A~0535+262 for 30 hours. 23 hours 40 minutes remain after the application of quality-selection criteria, which remove data taken during bad weather or 
affected by hardware-related problems. For some observations only three telescopes were operational (missing telescope 4 on 2009 December 24 and 2010 February 10), 
leading to a moderately reduced sensitivity during these periods. The triggering criterion (a counting rate of $>$0.1 counts s$^{-1}$ reported by the Swift/BAT instrument) for observations of flaring X-ray binaries with VERITAS was fulfilled on 2009 December 5 \citep{Wilson-Hodge-2010}. Observations started on 2009 December 6, shortly after the beginning of the giant flare. They were delayed by one day due to very bright moonlight conditions. The observation covered most of the flare, included the apastron phase, and continued for almost 90 days until the following periastron phase. Table \ref{table:Vdaily} shows the daily observation log, including weather condition, elevation range and moonlight condition. Figure \ref{fig:Slightcurve} (right) shows the relative orbit of 1A~0535+262 around the Be companion star.

The analysis steps consist of calibration, image cleaning, second-moment parameterization of the recorded images \citep{Hillas-1985}, reconstruction of shower direction and impact parameter using stereoscopic methods (see e.g.~\cite{Krawczynski-2006}), gamma-ray/hadron separation and the generation of sky maps. Images in at least three cameras are required; additional cuts on the shape of the event images and direction of the primary particles are used to reject the far more numerous background events. These cuts are optimized for a 5\% Crab Nebula flux source.
The energy threshold after analysis cuts is 220 GeV at a 10$^{\mathrm{o}}$ zenith angle and 450 GeV at a 40$^{\mathrm{o}}$ zenith angle and the systematic error on the energy estimation of the primary gamma rays is about 20\%. The data were taken in wobble mode \citep{Fomin-1994}, wherein the object was positioned at a fixed offset of 0.5$^{\mathrm{o}}$ in one of four directions from the camera center. The search region for photons from the putative gamma-ray source is defined by a 0.1$^{\mathrm{o}}$ radius circle centered on the position of the Be star in 1A~0535+262 \citep{Perryman-1997}. The background in the source region is estimated from the same field of view using the ring-background model with a ring size of 0.5$^{\mathrm{o}}$ (mean radius) and a ring width of 0.1$^{\mathrm{o}}$ \citep{Aharonian-2005b}. In order to reduce systematic errors in the background estimation, regions around stars with B magnitudes brighter than 6 are excluded.

\subsection{Fermi LAT: High-Energy Gamma-Ray Observations}
Fermi was placed in orbit on 2008 June 11 and has been acquiring data primarily in all-sky survey mode since 2008 August 4. It carries on board two science instruments: the \emph{Gamma-ray Burst Monitor} (GBM) and the \emph{Large Area Telescope} (LAT). For this work we only used data from the LAT, which is sensitive to high energy photons (HE; E$>$0.1 GeV). The LAT is an e$^{+}$e$^{-}$ pair production telescope containing solid-state silicon trackers and cesium iodide calorimeters, which are sensitive to photons in the energy range from $\sim$20 MeV to $\sim$300 GeV . It has a field of view of $\sim$2.4 sr, an on-axis effective area of $\sim$8000 cm$^2$ for energies $\gtrsim$1 GeV, and an angular resolution (for 68\% containment at 1 GeV) better than 1$^{\circ}$ \citep{Atwood-2009}. 

For this work we used data from LAT observations of 1A~0535+262 that were conducted between 2009 November 30 and 2010 February 22. We used the Fermi Science Tools v9r15p2 software analysis package to reduce the data and followed the event selection recommendations from the Fermi Science Support Center\footnote{\protect\url{http://fermi.gsfc.nasa.gov/ssc/data/analysis/scitools/}}. Briefly, we selected photons from the Pass 6 diffuse class events, those that have the highest probability of being gamma rays, with the {\tt gtselect} tool. In order to avoid contamination from Earth albedo photons, time periods when the region around 1A~0535+262 was observed at zenith angles greater than 105$^{\circ}$ were eliminated from further analysis.  We also limited the spectral range to above 200 MeV to reduce contamination by the galactic diffuse gamma-ray background.

At low energies, the point spread function (PSF) of the LAT is quite large, so it was necessary to deal with contamination by potential sources in the vicinity of 1A~0535+262. For this reason, a circular region of interest (RoI), centered on 1A~0535+262, was chosen with a radius of 8.5$^{\circ}$. We chose a source region of radius 17$^{\circ}$, also centered on 1A~0535+262, that encompasses the supernova remnant IC 443, which is 9.5$^{\circ}$ from 1A~0535+262, and the Geminga pulsar, which is 15.3$^{\circ}$ away. Both of these sources are known GeV-TeV gamma-ray emitters. IC 443 and Geminga were modeled as described by \citet{Abdo-2010a} and \citet{Abdo-2010b}, respectively.  All sources in the 11-month Fermi LAT catalog \citep{Abdo-2010c} inside the source region were considered. The catalog consists of data acquired from 2008 August 4 to 2009 July 4, which in the case of 1A~0535+262 contains data over several orbital periods but before the giant December 2009 outburst. There is no detection for 1A~0535+262 in the catalog. 1A~0535+262 is modeled assuming a fixed spectral index of 2.1 while the normalization is left as a free parameter. The sources were assumed to have a power-law spectrum, which was derived from the fluxes in the Fermi LAT catalog. 

Two differences from the 11-month catalog were encountered during the analysis. First, the source 1FGL~J0623.5+3330 possessed a test statistic (TS) value ten times higher than that reported in the Fermi catalog (see \citet{Mattox-1996} for further information on the test statistic). To account for such variability, we left both the normalization and spectral index free. Second, an unidentified source was also found at the position RA 74.12\arcdeg\ and Dec 26.99\arcdeg\, with a TS value of 68, and it was included in the source model. The spectral parameters were allowed to vary during the fit for sources that lie within the RoI but were fixed for those outside of the RoI (but still inside of the source region). An improvement to the source model fit was achieved when IC 443, which is in the source region, was allowed to vary with free model parameters. Besides discrete sources, contamination by the diffuse background was also taken into account. The galactic and extragalactic backgrounds were represented with \emph{gll\_iem\_v02.fit} and with a power-law model of index 4.09, respectively. The instrument response function (IRF) used in this analysis is IRF P6\_V3\_DIFFUSE.

%
%
%

\subsection{\label{sec:XRTanalysis}Swift/XRT: X-Ray Observations}

Swift is a gamma-ray burst satellite that contains three scientific instruments on board: the \emph{Burst Alert Telescope} (BAT), the \emph{X-ray Telescope} (XRT), and the \emph{UV/Optical Telescope} (UVOT). With a field of view of 2 sr, the BAT is an effective hard X-ray sky monitor. The detailed studies presented in this work relied on data from the XRT. The XRT operates in the 0.3-10 keV band. It consists of a focusing X-ray telescope and CCD detectors at the focal plane \citep{Burrows-2004}.

Triggered by alerts from the BAT, 1A~0535+262 was observed with the XRT between 2009 December 7 and December 27. Table \ref{table:XRTLog} provides a summary of the observations. Due to the brightness of the source, the Windowed Timing (WT) mode \citep{Capalbi-2005} was adopted to minimize the effects of photon pile-up. The XRT data were reduced and analyzed with FTOOLS in the HEASOFT package version 6.5.\footnote{\protect\url{http://heasarc.gsfc.nasa.gov/docs/software/lheasoft/}}  For each observation, initial event filtering and selection was carried out using the \texttt{xrtpipeline} script, with standard quality cuts, and only events with grades 0-2 were selected as good events. Source counts were extracted, with \texttt{xselect}, from a $20\times60$ pixel rectangular box centered on 1A~0535+262,  while background counts were taken from a rectangular region of the same size outside of the source region. An exposure map was generated with \texttt{xrtexpomap} and was used to correct for bad columns. Finally, source and background spectra were constructed. The spectra were grouped to contain a minimum of 100 counts per bin and a 1\% systematic error was added to the data \citep{Steiner-2010}. To facilitate subsequent spectral analyses, an ancillary response file (\texttt{arf}) was produced with \texttt{xrtmkarf}, to go along with the adopted response matrix (\textit{swxwt0to2s0\_20010101v011.rmf}). 

Even in the Windowed Mode, event pile-up was significant in some observations. For instance,  near the peak of the outburst (on 2009 December 8), the pile-up of photons reached about 11\%. Therefore, the effects need to be properly accounted for. We adopted a procedure similar to that of \citet{Romano-2006} to correct the pile-up effects.  Briefly, for each observation, we constructed an X-ray spectrum by excluding a central region of the source.  We varied the size of the exclusion box until the spectrum was stabilized (based on the results of spectral fitting). After the procedure, the remaining pile-up effects on the spectrum are expected to be minimal.

\subsection{RXTE/PCA: X-Ray Observations}

Besides the ASM, RXTE carries two pointed instruments: the \emph{Proportional Counter Array} (PCA) and the \emph{High Energy X-ray Timing Experiment} (HEXTE). For this work only data from the PCA were employed. The PCA was designed to cover an energy range of 2.0--60.0 keV and consists of five, nearly identical, large-area xenon proportional counter units (PCUs). The field of view is about $1^{\circ}\!\times1^{\circ}$. Each PCU has an area of $\sim$1300 cm$^{2}$, for a total effective area of 6500 cm$^{2}$ \citep{Jahoda-1996}. Operational constraints often require that some of the PCUs be turned off, but PCU 2 was always in operation for our observations during the outburst.

1A~0535+262 was observed by RXTE between 2009 December 4 and 2010 January 4. Table \ref{table:RXTELog} shows an observation log. We reduced the data using FTOOLS in the HEASOFT package (version 6.5)\footnote{\protect\url{http://heasarc.gsfc.nasa.gov/docs/software/lheasoft/}}. The PCA has numerous data modes, and multiple modes are usually employed in an observation. For this work, however, we only used the {\it Standard 2} data. For a given observation, we first filtered data by following the standard procedure\footnote{See the online RXTE Cook Book at http://heasarc.gsfc.nasa.gov/docs/xte/recipes/cook\_book.html},  which resulted in a list of good time intervals (GTIs). We then simulated background events for the observation by using the latest background model (\textit{pca\_bkgd\_cmbrightvle\_eMv20051128.mdl}) that is appropriate for bright sources. Using the GTIs, we proceeded to make a spectrum for PCU~2 by using data from only the first xenon layer (which is most accurately calibrated), which limits the spectral coverage to below 25--30 keV. Since few counts were detected at higher energies, the impact of the reduced spectral coverage was very minimal. The selected energy range is below the range where 1A~0535+262 produces cyclotron lines, which are due to the first and second harmonic cyclotron absorption at $\sim\!45$ keV and $\sim\!100$ keV \citep{Kendziorra-1994,Caballero-2007b}. We repeated the steps to derive a corresponding background spectrum for PCU 2 from the simulated events. We added 1\% systematic uncertainty to the spectra for subsequent spectral modeling to account for the uncertainty in the instrumental response \citep{Jahoda-2006}.
 
\section{Results}

\subsection{VERITAS}
\begin{figure}
\plotone{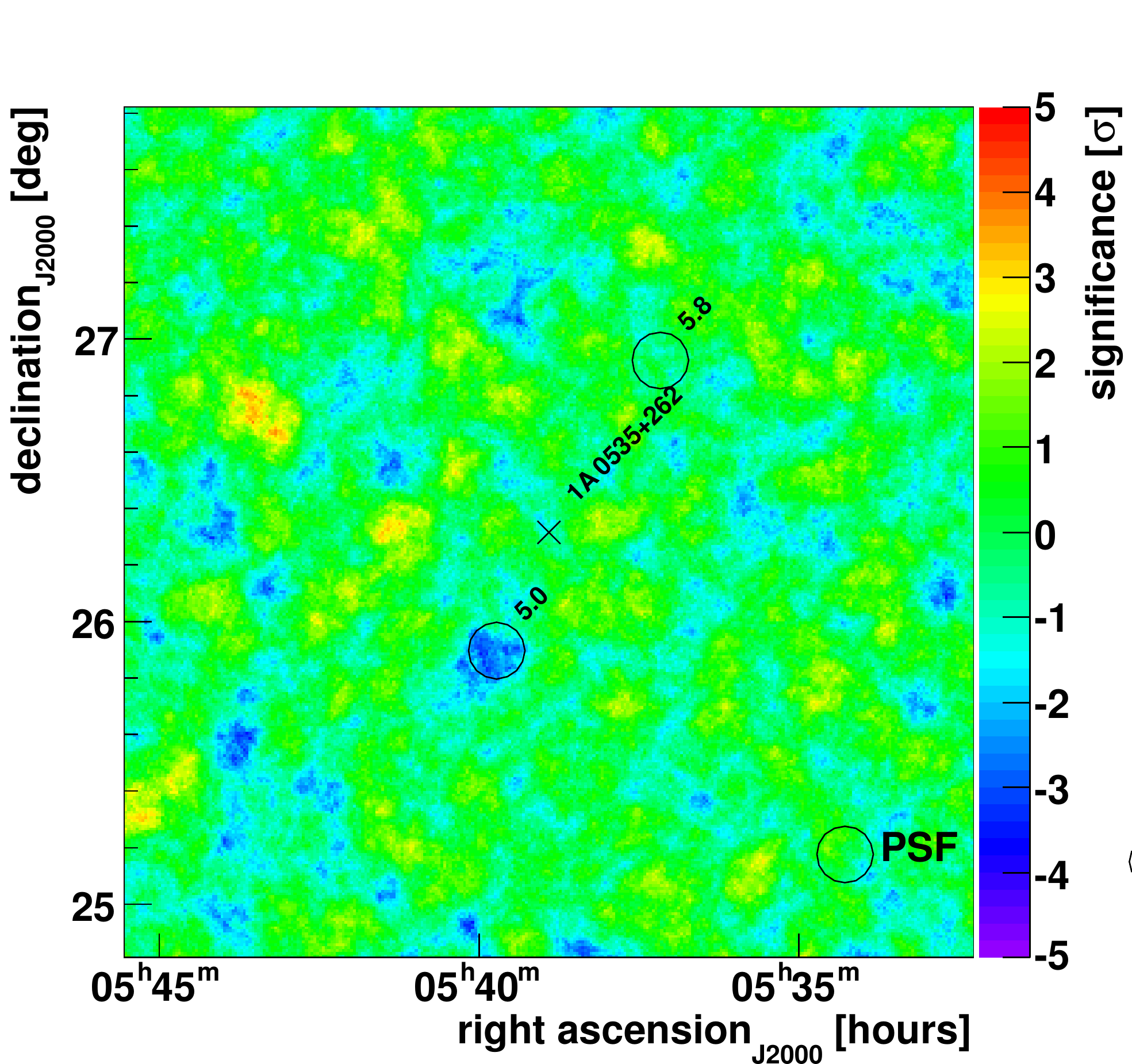}                                                                                          
\caption{\label{fig:Vskyplot}
VERITAS significance map of the region around 1A~0535+262 in equatorial coordinates
for the period MJD 57166-55250. The location of 1A~0535+262 is indicated by a black cross at the center. Also shown are regions excluded from the background calculation due to bright stars. The numbers beside the excluded regions indicate the B magnitude of these stars.  The circle at the bottom right indicates the angular resolution of the VERITAS observations.
}
\end{figure}

Analysis results for the combined VERITAS data set can be found in Table \ref{table:Vresults}. Figure \ref{fig:Vskyplot} shows the region of the sky around 1A~0535+262 as seen by VERITAS. No evidence for VHE gamma rays has been found.
The flux upper limit at the 99\% confidence level \citep{Helene-1983} assuming a power-law-like source spectrum with a spectral index of $\Gamma=2.5$ is $F(>0.3$ TeV$) < 0.5 \times 10^{-12}$ ph cm$^{-2}$ s$^{-1}$ (about 0.4\% of the flux of the Crab Nebula above 0.3 TeV).

The data were arranged in different periods, as gamma-ray production and absorption is expected to vary with orbital movement and flaring state. The four periods are: rising portion and falling portion of the giant flare, apastron and periastron. No VHE emission has been detected in any of these periods; upper limits between 0.9 and 2.0\% of the flux of the Crab Nebula ($>$ 0.3 TeV) have been derived. For details, see Table \ref{table:Vresults}.

\subsection{Fermi}
A search for HE gamma-ray emission from 1A~0535+262 was performed for a period that spans the onset of the X-ray outburst to the successive apastron of the pulsar (2009 November 30 -- 2010 February 22). No significant gamma-ray excess was seen for the same time intervals used for the VERITAS data analysis. We derived a flux upper limit of $F(>0.2$ GeV)$<1.9\times10^{-8}$ ph cm$^{-2}$ s$^{-1}$ at the 99\% confidence level. To facilitate a comparison with theoretical models, we also derived flux upper limits 
for different periods of the X-ray outburst (rising and decaying portions) and around apastron and periastron passages. The results are summarized in Table \ref{table:Fresults}.

\subsection{Swift/XRT}
The XRT spectrum for each time interval was modeled in XSPEC version 12.6.0k. We used response matrix file (\texttt{rmf}) v011, along with the \texttt{arf} that we made. During spectral fitting, we limited the energy range to 0.6-10.0 keV, to avoid complicated calibration issues at the lower energies. We should note the presence of two absorption features, at 1.8 and 2.2 keV, respectively, which are  probably of instrumental origin as they lie near the Si K and Au M edges.\footnote{See~\protect\url{http://heasarc.gsfc.nasa.gov/docs/heasarc/caldb/swift/docs/xrt/SWIFT-XRT-CALDB-09\_v14.pdf}}.  They were modeled with inverted Gaussians at fixed energies in the spectral modeling. 

We fitted the spectra with a model that consists of a blackbody and a power law, both of which are absorbed by foreground matter in the interstellar medium. Good fits were achieved in all cases. The results are shown in Table~\ref{table:XRTFits}. Spectral hardening during the rising phase of the X-ray outburst is apparent.  The reverse trend is noticeable during the decaying phase. We note that the rising phase corresponds to orbital phases 0.11--0.21 and the decaying phase to orbital phases 0.22--0.36. 

\subsection{RXTE/PCA}
Similarly, we modeled the PCA spectra in XSPEC. We limited the energy range to $>$3 keV, to avoid calibration issues at lower energies. Due to the lack of sensitivity at low energies, for each observation we fixed the hydrogen column density $N_H$ to $3.0\times10^{21}$ cm$^{-2}$, which is fairly representative of the values derived from the Swift/XRT observation (see Table~\ref{table:XRTFits}). We found that the simple model that had worked well for the Swift/XRT spectra could no longer satisfactorily fit the PCA spectra. Examining the residuals, we noted the presence of a broad feature between 6 and 7 keV, which  
is likely caused by calibration uncertainty near the Xe L edge, although it could also be partially attributable to a K$_{\alpha}$ line of neutral or ionized iron. 
We modeled it with a Gaussian and will not discuss it further. More interestingly, we found that we could achieve good fits to the PCA spectra by introducing a high-energy cut-off.  

The roll-over of the spectrum at high energies seems to suggest a physical origin of the hard X-ray emission in thermal inverse Compton scattering. The broad spectral coverage of the RXTE/PCA allows modeling of the X-ray spectrum with a more physical model. We replaced the empirical power law (with a high-energy cut-off) with the {\it comptt} model in XSPEC.  Table \ref{table:RXTEFits} summarizes the results. The fits are all excellent. Due to the lack of sensitivity of the PCA at low energies, however,  the seed photon temperatures ($kT_{s}$) is not well constrained. This also affects other parameters, such as the normalization of {\it comptt}, because of the coupling between the spectral components.

\subsection{\label{sec:Swift-RXTE}Joint Swift/RXTE Analysis}
Ideally we would combine both the Swift/XRT and RXTE/PCA data to constrain all spectral components. Unfortunately there were no simultaneous observations, but there were many sets of observations from both satellites which occurred within the timescale of a day. The average energy flux would vary around $\sim$1 keV cm$^{-2}$ s$^{-1}$ per day during the maximum X-ray outburst. We attempted to model these pairs of Swift/XRT and RXTE/PCA observations jointly, with the understanding that the source could have varied spectrally between the two. To account for possible discrepancy in the overall throughput between the XRT and PCA, we introduced a multiplicative constant to the models. We fixed the constant to unity for the XRT but let it float for the PCA.

\begin{figure}
\plottwo{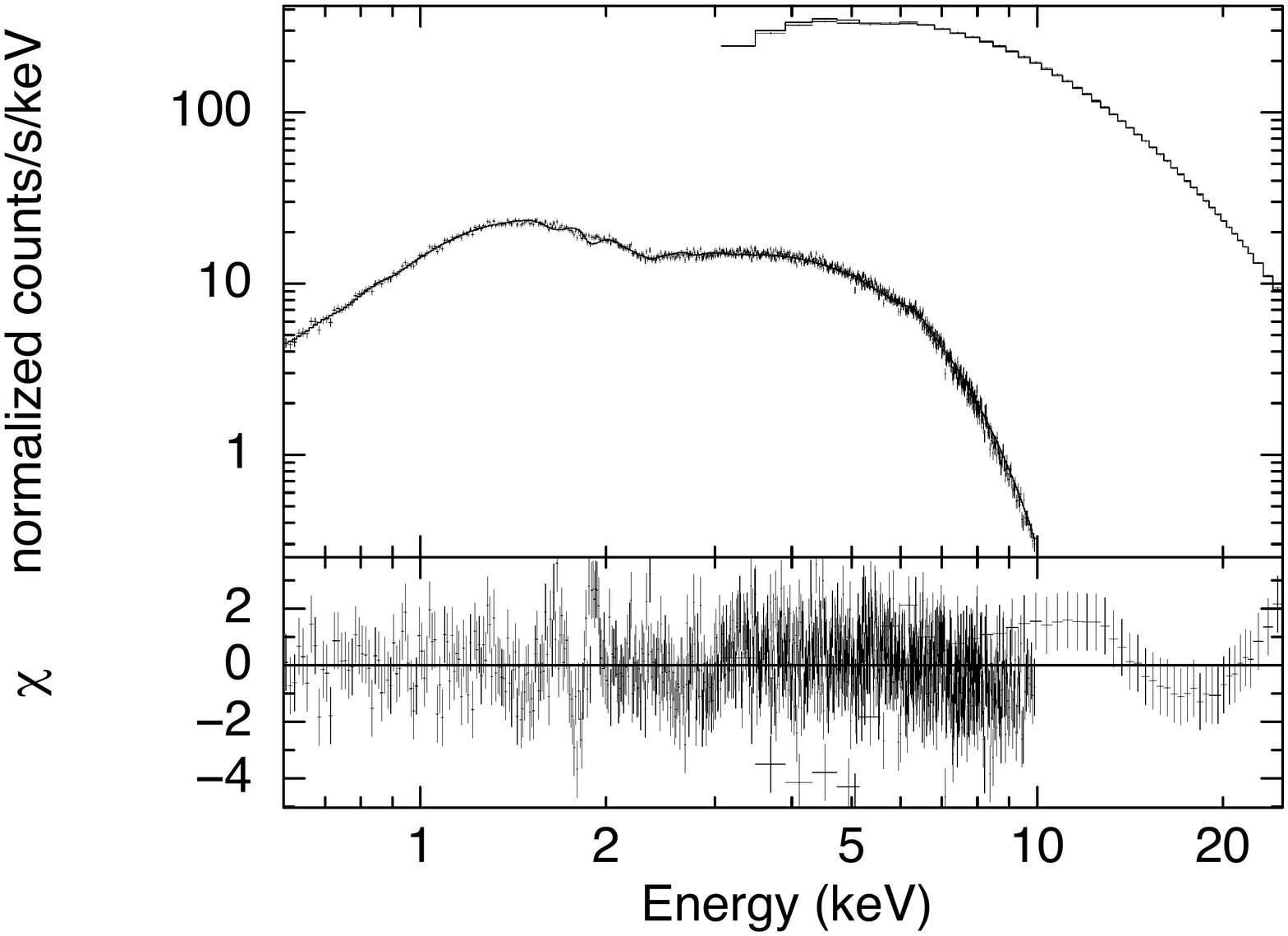}{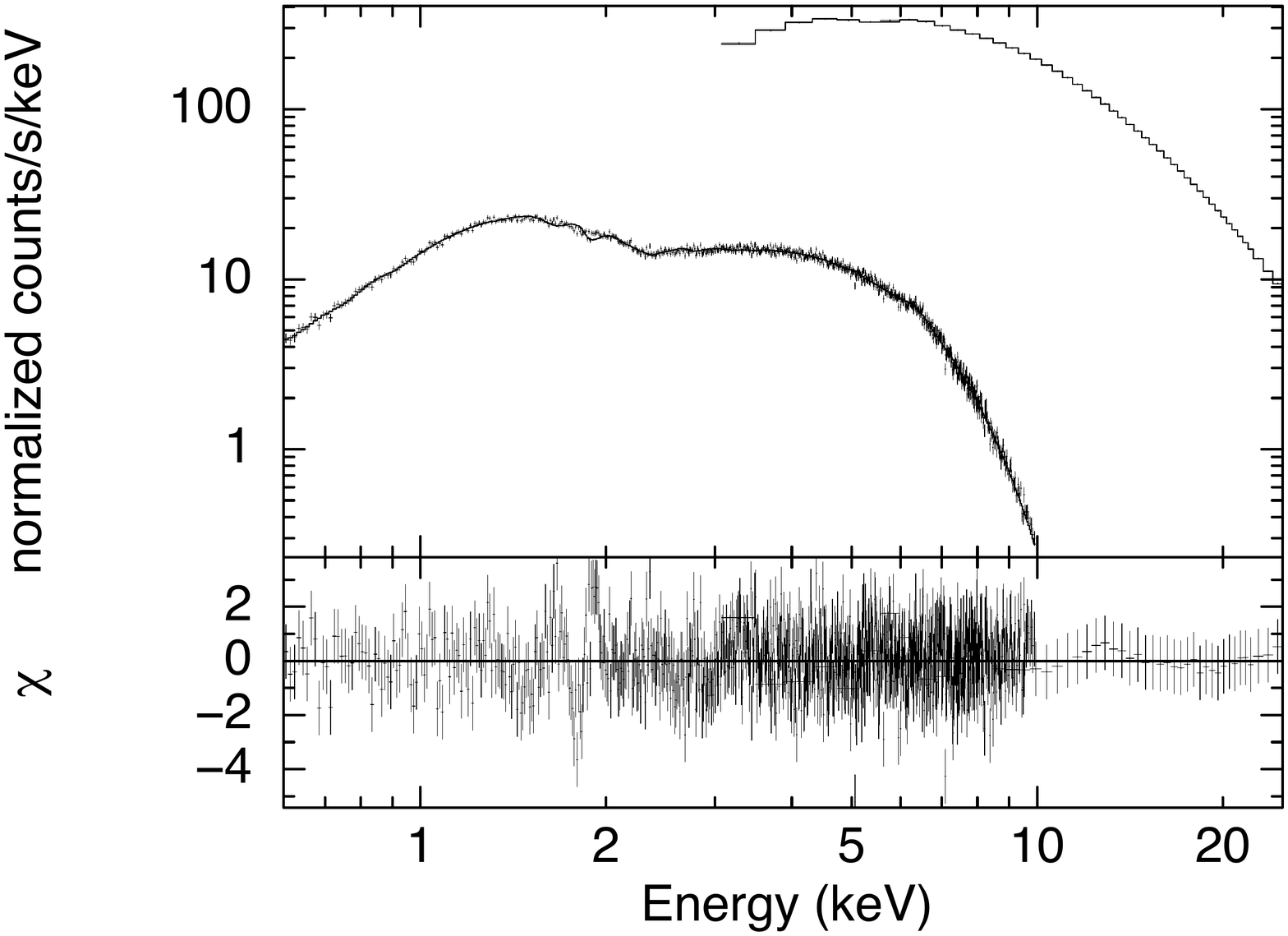}
\caption{\label{fig:comp}
Joint Swift/XRT and RXTE/PCA spectral modeling of 1A~0535+262 for the rising portion of the X-ray outburst. The upper panel shows the Swift/XRT data (lower curve) and the RXTE/PCA data (upper curve), and the lower panel shows the residuals of the fit. ({\it left}) A fit to the data with a model that consists of blackbody radiation and thermal Comptonization, with all physical parameters tied for the XRT and PCA data sets. Note a significant pattern in the residuals of the fit (shown in the bottom panel) in the PCA band. ({\it right}) A fit to the data with the same model but with the blackbody temperature and normalization untied between the two data sets.
}
\end{figure}

We experimented with several physical models to gain further insights into the origin of  the observed spectral roll-over at high energies, including thermal Comptonization ({\it comptt}), thermal Bremsstrahlung ({\it bremss}), and non-thermal synchrotron ({\it srcut}), in combination with other spectral components that were used in the modeling of the Swift/XRT and RXTE/PCA data (mainly at lower energies). We were not able to find any acceptable fits when we tied all of the physical parameters in the model for the Swift/XRT and RXTE/PCA observations, but when we untied the blackbody parameters (i.e., let them vary independently for the Swift/XRT and RXTE/PCA observations), we obtained a good fit only with the thermal Comptonization model (for hard X-ray emission). This is perhaps an indication that the source did indeed vary significantly between the two observations and seems to indicate that the hard X rays are likely of thermal origin. Figure~\ref{fig:comp} shows an example (for the rising portion; see Table~\ref{table:JointSwift/RXTE}) of the fits and residuals for cases where the blackbody parameters are tied and untied, respectively. The results are summarized in Table~\ref{table:JointSwift/RXTE}.

\section{Summary and Discussion}

We present observations at X-ray and gamma-ray energies of the Be/pulsar binary 1A~0535+262 during its 2009 giant outburst. The results can be summarized as follows: 
i) There is no evidence for VHE or HE emission from 1A~0535+262 in the VERITAS and Fermi/LAT observations during the giant outburst and during the subsequent apastron and periastron passages; 
ii) The X-ray spectra measured with the Swift/XRT and RXTE/PCA are best fitted with a model that consists of blackbody and Comptonized emission from thermal electrons at temperatures of approximately 2 keV and 6 keV,  respectively. The optically-thick emission may originate from the accretion disk around the neutron star and perhaps also from the ``hot spot'' on the neutron star surface. The optical depths for the Compton component are $\sim$10 (see Table~\ref{table:Swift-RXTE}). This emission component may be associated with, e.g.,  a warm layer of the accretion disk or with the accretion column above the ``hot spot''. 

The non-detection of 1A~0535+262 at VHE and HE wavelengths may indicate that there is no significant non-thermal particle population in the system. This would imply a thermal origin of the X-ray emission, as opposed to a non-thermal leptonic model, in which X-ray emission is the result of synchrotron radiation from non-thermal electrons. A thermal origin is also supported by the fact that 1A~0535+262 has not been detected at radio wavelengths \citep{Giovanelli-1992, Romero-2001}, which also suggests the lack of non-thermal electrons, as well as by the OSSE/CGRO observation of the source that saw no significant non-thermal component~\citep{Grove-1995}. 

The upper limits derived from VERITAS observations correspond to a luminosity of $<\!0.5-1.5\times 10^{33}$ erg s$^{-1}$ (see Table \ref{table:Vresults}), assuming a distance to 1A~0535+262 of 2 kpc. 
Applying the Cheng and Ruderman mechanism to the source, Orellana et al.  (2007) derived a gamma-ray luminosity of about $10^{33}$ erg s$^{-1}$ at 0.3 TeV at the end of giant outbursts~\citep{Orellana-2007}, which is very close to our upper limits.  The Fermi/LAT flux upper limit over the whole orbit (see Table \ref{table:Fresults}) is already below the theoretical flux prediction of $3.8\times10^{-8}$ ph cm$^{-2}$ s$^{-1}$, which was derived by extrapolating the result of \citet{Orellana-2007} to the Fermi/LAT energy range. Therefore, our results begin to place severe constraints on hadronic models as well. We note that the upper limits correspond only to a tiny fraction of the Eddington luminosity of the system. In comparison, the X-ray luminosity of 1A~0535+262 can reach about 10\% of the Eddington luminosity.

One of the main interests in 1A~0535+262 stems from the previous EGRET gamma-ray source detection before the February 1994 major X-ray outburst peak. There is no Fermi detection for the EGRET source 3EG~J0542+2160, and a 99\% confidence level flux upper limit at the source location  (RA 85.69$^{\circ}$, Dec 26.17$^{\circ}$) produces a value of F($>0.2$GeV)$<3.5\times 10^{-8}$ ph cm$^2$ s$^{-1}$ (test statistic TS is 1.0), for the same source model settings and energy range used for 1A~0535+262 over the whole orbital period. 1FGL~J0538.6+2717, the closest catalog source to 1A~0535+262, is 0.98$^{\circ}$ away and doesn't overlap with the 95\% confidence level location radius of 3EG J0542+2610. 

The lack of detectable gamma-ray emission may also be attributed to the attenuation of gamma rays via pair production, because of the presence of a strong radiation field in the binary system at both optical and X-ray wavelengths. The VHE gamma rays should be attenuated mainly by IR photons from the companion (Oe) star and are thus expected to be modulated by the orbital motion~~\citep{Bednarek-2006}. However, quantifying the effects of attenuation is complicated by dramatic changes in the accretion rate that are associated with the giant outburst, since the accretion process could also be a source of IR photons and, more importantly, could power the VHE gamma-ray production. In other words, there is significant degeneracy in the production and attenuation of VHE gamma rays. It is worth noting that the secondary electrons (and positrons) from the pair production process could be a source of gamma rays at MeV--GeV energies \citep{Bednarek-2006}. The HE gamma rays can be similarly attenuated, mainly by soft X-ray photons. We therefore do not expect to detect such gamma rays near the peak of the X-ray outburst. However, the fact that 1A~0535+262 is not detected with the Fermi/LAT when the attenuation of HE photons is not expected to be significant seems to indicate a genuine lack of gamma-ray production. 

Our results seem to suggest that 1A~0535+262 is inherently different from those Be X-ray binaries that have been detected at GeV--TeV energies, including PSR~B1259--63 and LS~I~+61$^\circ$303, in terms of gamma-ray production.
Observationally, these systems tend to be radio sources and their X-ray spectra contain a significant non-thermal component.
Physically, while the nature of the compact object in LS~I~+61$^\circ$303 is still uncertain, PSR~B1259--63 contains a rapidly rotating pulsar with a much lower spin period than 1A~0535+262. These two systems present more extreme physical conditions than 1A~0535+262, but equally extreme physical conditions exists in systems which are also undetected in gamma rays. The environmental conditions which lead to gamma-ray production in binary systems remain poorly defined. To more meaningfully constrain theoretical models on gamma-ray production in 1A~0535+262, we would probably need to lower the VHE gamma-ray upper limits by an order of magnitude. This source represents the archetype of the class of Be binary systems which exhibit giant outbursts and the VHE observations were the best we could hope to get in terms of coverage and exposure. The source will be a good target for the next-generation ground-based gamma-ray observatories.

\acknowledgments

This research is supported by grants from the U.S. Department of Energy, the U.S. National Science Foundation, the Smithsonian Institution, by NSERC in Canada, by Science Foundation Ireland, and by STFC in the UK. We acknowledge the excellent work of the technical support staff at the FLWO and the collaborating institutions in the construction and operation of the instrument. GM acknowledges support through the Young Investigators Program of the Helmholtz Association. AV and WC wish to acknowledge financial support from NASA and Purdue University.

{\it Facilities:} \facility{VERITAS}, \facility{Swift}, \facility{RXTE}, \facility{Fermi LAT}

\clearpage


\begin{deluxetable}{ccccc}
\tabletypesize{\scriptsize}
\centering
\tablecolumns{2}
\tablewidth{0pt}
\tablecaption{VERITAS observation log
\label{table:Vdaily}}
\tablehead{
\colhead{Date} &
\colhead{Observation} &
\colhead{Elevation} &
\colhead{N$_{\mathrm{tel}}$} &
\colhead{Observing} \\
\colhead{} &
\colhead{Time (min)} &
\colhead{Range} &
\colhead{} &
\colhead{Conditions} 
}
\startdata
2009/12/06 & 20 & $78^{\mathrm{o}}-80^{\mathrm{o}}$ & 4 & moon 80\% illuminated  \\
2009/12/07 & 40 & $58^{\mathrm{o}}-65^{\mathrm{o}}$ & 4 & bad weather, moon 70\% illuminated  \\
2009/12/09 & 38 & $83^{\mathrm{o}}-85^{\mathrm{o}}$ & 4 & moon 48\% illuminated  \\
2009/12/10 & 176 & $48^{\mathrm{o}}-84^{\mathrm{o}}$ & 4 &  moon 37\% illuminated  \\
2009/12/11 & 52 & $56^{\mathrm{o}}-80^{\mathrm{o}}$ & 4 & -  \\
2009/12/12 & 40 & $62^{\mathrm{o}}-65^{\mathrm{o}}$ & 4 & -  \\
2009/12/15 & 40 & $53^{\mathrm{o}}-56^{\mathrm{o}}$ & 4 & -  \\
2009/12/16 & 60 & $65^{\mathrm{o}}-74^{\mathrm{o}}$ & 4 & -  \\
2009/12/17 & 60 & $68^{\mathrm{o}}-75^{\mathrm{o}}$ & 4 & -  \\
2009/12/18 & 60 & $75^{\mathrm{o}}-82^{\mathrm{o}}$ & 4 & -  \\
2009/12/19 & 60 & $61^{\mathrm{o}}-69^{\mathrm{o}}$ & 4 & -  \\
2009/12/20 & 60 & $61^{\mathrm{o}}-70^{\mathrm{o}}$ & 4 & -  \\
2009/12/21 & 80 & $67^{\mathrm{o}}-81^{\mathrm{o}}$ & 4 & bad weather \\
2009/12/24 & 60 & $59^{\mathrm{o}}-68^{\mathrm{o}}$ & 3 & -  \\
2009/12/25 & 60 & $49^{\mathrm{o}}-58^{\mathrm{o}}$ & 4 & -  \\
2009/12/26 & 40 & $55^{\mathrm{o}}-57^{\mathrm{o}}$ & 4 & - \\
2010/01/05 & 37 & $76^{\mathrm{o}}-80^{\mathrm{o}}$ & 4 & -  \\
2010/01/07 & 60 & $69^{\mathrm{o}}-76^{\mathrm{o}}$ & 4 & -  \\
2010/01/09 & 43 & $80^{\mathrm{o}}-85^{\mathrm{o}}$ & 4 & bad weather \\
2010/01/11 & 8   & $68^{\mathrm{o}}-70^{\mathrm{o}}$ & 4 & -  \\
2010/01/12 & 60 & $70^{\mathrm{o}}-78^{\mathrm{o}}$ & 4 & -  \\
2010/01/14 & 38 & $71^{\mathrm{o}}-76^{\mathrm{o}}$ & 4 & -  \\
2010/01/16 & 60 & $73^{\mathrm{o}}-81^{\mathrm{o}}$ & 4 & -  \\
2010/01/17 & 60 & $57^{\mathrm{o}}-65^{\mathrm{o}}$ & 4 & -  \\
2010/02/05 & 12 & $82^{\mathrm{o}}-83^{\mathrm{o}}$ & 4 & -  \\
2010/02/07 & 20 & $77^{\mathrm{o}}-80^{\mathrm{o}}$ & 4 & -  \\
2010/02/08 & 40 & $54^{\mathrm{o}}-62^{\mathrm{o}}$ & 4 & - \\
2010/02/10 & 37 & $60^{\mathrm{o}}-71^{\mathrm{o}}$ & 3/4 & -  \\
2010/02/12 & 60 & $79^{\mathrm{o}}-84^{\mathrm{o}}$ & 4 & -  \\
2010/02/16 & 60 & $70^{\mathrm{o}}-79^{\mathrm{o}}$ & 4 & -  \\
2010/02/18 & 60 & $72^{\mathrm{o}}-82^{\mathrm{o}}$ & 4 & moon 89\% illuminated  \\
2010/02/20 & 60 & $75^{\mathrm{o}}-83^{\mathrm{o}}$ & 4 & moon 98\% illuminated  \\
\enddata
\tablecomments{Data taken in bad weather or under very bright moonlight conditions ($>50$\% illumination) have been excluded from the data analysis.
The column N$_{\mathrm{tel}}$ contains the number of working telescopes, from 3 to 4.}
\end{deluxetable}

\begin{deluxetable}{ccccc}
\tabletypesize{\scriptsize}
\centering
\tablecolumns{5}
\tablewidth{0pt}
\tablecaption{\label{table:XRTLog}Swift/XRT observation log}
\tablehead{
\colhead{Obs ID} &
\colhead{Start} &
\colhead{End} &
\colhead{Exposure Time} &
\colhead{Count Rate\tablenotemark{a}} \\
\colhead{} &
\colhead{(UT; 2009)} &
\colhead{(UT; 2009)} &
\colhead{(sec)} &
\colhead{(cts s$^{-1}$)} 
}
\startdata
\sidehead{Rising portion of X-ray flare (MJD 55166.4 -- 55177.6)}
00035066002 & 2009-12-07 00:25:13 & 2009-12-08 00:30:00 & 981 & 195 \\
00035066005 & 2009-12-08 22:46:41 & 2009-12-08 23:03:21 & 995 & 227 \\
00035066006 & 2009-12-09 16:08:26 & 2009-12-09 16:28:00 & 1154 & 239 \\ 
00035066007 & 2009-12-10 10:09:54 & 2009-12-10 16:28:00 & 983 & 234 \\
00035066008 & 2009-12-11 00:33:44 & 2009-12-11 00:51:00 & 985 & 246 \\
00035066009 & 2009-12-12 00:42:28 & 2009-12-12 00:59:00 & 977 & 151 \\
\hline \\
\sidehead{Falling portion of X-ray flare (MJD 55178.4 -- 55193.6)}
00035066010 & 2009-12-13 00:46:28 & 2009-12-13 01:03:00 & 972 & 228  \\
00035066011 & 2009-12-14 00:47:31 & 2009-12-14 01:04:00 & 962 & 242 \\
00035066012 & 2009-12-15 00:52:43 & 2009-12-15 01:11:00 & 1083 & 242 \\
00035066013 & 2009-12-16 00:40:25 & 2009-12-16 00:58:00 & 1050 & 240 \\
00035066014 & 2009-12-18 00:50:32 & 2009-12-18 01:07:00 & 943 & 200 \\
00035066015 & 2009-12-19 01:16:47 & 2009-12-19 01:33:00 & 968 & 177 \\
00035066017 & 2009-12-21 01:11:21 & 2009-12-21 01:28:00 & 992 & 182 \\
00035066018 & 2009-12-22 09:13:48 & 2009-12-22 09:35:00 & 1252 & 163 \\
00035066020 & 2009-12-23 09:26:21 & 2009-12-23 09:45:00 & 1105 & 145 \\
00035066021 & 2009-12-24 09:24:36 & 2009-12-24 09:42:00 & 1029 & 133 \\
00035066022 & 2009-12-25 08:11:28 & 2009-12-25 08:31:00 & 1123 & 98 \\
00035066023 & 2009-12-26 08:14:49 & 2009-12-26 08:35:00 & 1199 & 109 \\
00035066024 & 2009-12-27 08:21:56 & 2009-12-27 08:41:00 & 1137 & 103 
\enddata
\tablenotetext{a}{Background-subtracted count rates in the 0.6--10.0 keV band. They are not pile-up corrected.}
\end{deluxetable}

\begin{deluxetable}{cccc}
\tabletypesize{\scriptsize}
\centering
\tablecolumns{4}
\tablewidth{0pt}
\tablecaption{\label{table:RXTELog}RXTE/PCA observation log}
\tablehead{
\colhead{Event ID} & 
\colhead{Start Time} &
\colhead{Stop Time} &
\colhead{Exposure Time} \\
\colhead{} &
\colhead{(UT)} &
\colhead{(UT)} &
\colhead{(sec)}
}
\startdata
\sidehead{Rising portion of X-ray flare (MJD 55166.4 -- 55177.6)}
94323-04-01-01 & 2009-12-04 10:27:12 & 2009-12-04 11:23:44 & 2448 \\
94323-04-01-00 & 2009-12-05 16:14:24 & 2009-12-05 17:58:40 & 3152 \\
94323-04-01-02 & 2009-12-06 12:36:32 & 2009-12-06 13:46:56 & 3200 \\ 
94323-04-01-03 & 2009-12-07 13:45:20 & 2009-12-07 14:57:36 & 3232 \\
94323-04-01-04 & 2009-12-08 14:45:36 & 2009-12-08 15:55:28 & 2896 \\
94323-04-01-05 & 2009-12-09 17:25:20 & 2009-12-09 18:48:32 & 3120 \\
94323-04-01-06 & 2009-12-10 16:55:12 & 2009-12-10 21:27:28 & 9952 \\
94323-04-02-00 & 2009-12-11 11:57:20 & 2009-12-11 12:59:28 & 2832 \\
94323-04-02-01 & 2009-12-12 13:04:16 & 2009-12-12 14:33:36 & 2944 \\
\hline \\
\sidehead{Falling portion of X-ray flare (MJD 55178.4 -- 55193.6)}
94323-04-02-02 & 2009-12-13 07:55:12 & 2009-12-13 08:42:40 & 2064 \\
94323-04-02-04 & 2009-12-14 18:29:20 & 2009-12-14 19:41:36 & 3376 \\
94323-05-01-00 & 2009-12-15 14:36:16 & 2009-12-15 16:05:36 & 3392 \\
94323-05-01-01 & 2009-12-16 11:22:24 & 2009-12-16 12:28:32 & 3328 \\
94323-05-01-02 & 2009-12-17 15:14:24 & 2009-12-17 16:46:40 & 3392 \\
94323-05-02-03 & 2009-12-18 14:50:24 & 2009-12-18 16:30:40 & 3328 \\
94323-05-02-00 & 2009-12-19 09:22:24 & 2009-12-19 13:03:28 & 6864 \\
94323-05-02-06 & 2009-12-19 14:25:20 & 2009-12-19 17:27:28 & 6720 \\
94323-05-02-04 & 2009-12-20 12:26:24 & 2009-12-20 15:35:28 & 6624 \\
94323-05-02-01 & 2009-12-21 07:46:24 & 2009-12-21 10:25:36 & 4528 \\
94323-05-02-02 & 2009-12-23 11:03:12 & 2009-12-23 15:54:40 & 9920 \\
94323-05-02-05 & 2009-12-24 10:59:40 & 2009-12-24 14:50:40 & 7664 \\
94323-05-03-00 & 2009-12-25 15:27:28 & 2009-12-25 16:32:48 & 2704 \\
94323-05-03-01 & 2009-12-26 11:26:24 & 2009-12-26 16:03:44 & 9072 \\
94323-05-03-02 & 2009-12-27 14:29:36 & 2009-12-27 15:34:40 & 9920 \\
94323-05-03-03 & 2009-12-28 17:19:28 & 2009-12-28 19:32:32 & 3328 \\
\hline \\
\sidehead{Apastron (MJD 55199.4 -- 55216.6)}
94323-05-04-00 & 2010-01-02 01:51:12 & 2010-01-02 03:31:28 & 2832 \\
94323-05-04-01 & 2010-01-04 00:58:24 & 2010-01-04 02:15:28 & 3136 \\
94323-05-04-02 & 2010-01-07 02:36:16 & 2010-01-07 03:56:32 & 3008 \\
\enddata
\end{deluxetable}

\begin{deluxetable}{ccccccccccc}
\tabletypesize{\scriptsize}
\centering
\tablecolumns{10}
\tablewidth{0pt}
\tablecaption{VERITAS Analysis Results
\label{table:Vresults}}
\tablehead{
\colhead{Period} &
\colhead{MJD} &
\colhead{Observation} &
\colhead{Elevation} &
\colhead{On} &
\colhead{Off} &
\colhead{alpha} &
\colhead{Excess} &
\colhead{Significance} &
\colhead{Flux Upper} \\
\colhead{} &
\colhead{Range} &
\colhead{Time} &
\colhead{Range} &
\colhead{Events} &
\colhead{Events} &
\colhead{} &
\colhead{Events} &
\colhead{($\sigma$)} &
\colhead{Limit} \\
\colhead{} &
\colhead{} &
\colhead{(min)} &
\colhead{} &
\colhead{} &
\colhead{} &
\colhead{} &
\colhead{} &
\colhead{} &
\colhead{($10^{-12}$ cm$^{-2}$ s$^{-1})$}
}
\startdata
All & 57166 -- 55249  & 1420 & $48^{\mathrm{o}}-85^{\mathrm{o}}$ & 86 & 801 & 0.13 &  -15.8 & -1.5 & 0.5 \\
\hline
Rising portion & 55166 -- 55178 & 305 & $48^{\mathrm{o}}-85^{\mathrm{o}}$  & 19 & 184 & 0.13 & -4.5 & -0.9 & 1.3 \\
Falling portion & 55178 -- 55193 & 501 & $49^{\mathrm{o}}-78^{\mathrm{o}}$  & 33 & 343 & 0.13 &  -10.7 & -1.6 & 0.9 \\
Apastron & 55199 -- 55216 & 323 & $57^{\mathrm{o}}-81^{\mathrm{o}}$  &  16 & 161 & 0.13 & -4.5 & -1.0 & 1.0 \\
Periastron & 55230 -- 55249 & 289 & $54^{\mathrm{o}}-85^{\mathrm{o}}$  & 18 & 113 & 0.13 & 3.7 & 0.9 & 2.0 \\
\enddata
\tablecomments{Upper limits (E$>0.3$ TeV) are given at 99\% confidence level (after \cite{Helene-1983}). Significances are calculated using equation (17) from \cite{Li-1983}.
}
\end{deluxetable}

\begin{deluxetable}{ccccc}
\tabletypesize{\scriptsize}
\centering
\tablecolumns{5}
\tablewidth{0pt}
\tablecaption{Fermi LAT Analysis Results
\label{table:Fresults}}
\tablehead{
\colhead{Period} &
\colhead{MJD} &
\colhead{Exposure Time} &
\colhead{Significance\tablenotemark{a}} &
\colhead{Flux Upper Limit\tablenotemark{b}} \\
\colhead{} &
\colhead{Range} &
\colhead{(sec)} &
\colhead{(TS)} &
\colhead{($10^{-8}$ cm$^{-2}$ s$^{-1})$}
}
\startdata
All & 55165.9 -- 55249.1 & 3125089 & 0.0 & 1.9 \\ 
\hline \\
Rising portion of X-ray flare & 55165.9 -- 55177.6 & 429518 & 0.0 & 6.4 \\ 
Falling portion of X-ray flare & 55178.4 -- 55193.6 & 515688 & 0.3 & 9.8 \\ 
Apastron & 55199.4 -- 55216.6 & 725252 & 0.0 & 6.8 \\ 
Periastron & 55230.4 -- 55249.6 & 667580 & 0.0 & 5.0 \\ 
\enddata
\tablenotetext{a}{The definition of the test statistic is given by equation (20) of \citet{Mattox-1996}.}
\tablenotetext{b}{The upper limits are given at the 99\% confidence level, for photon energies above 0.2 GeV.
}
\end{deluxetable}

\begin{deluxetable}{ccccccc}
\tabletypesize{\scriptsize}
\centering
\tablecolumns{7}
\tablewidth{0pt}
\tablecaption{\label{table:XRTFits}Swift/XRT Spectral Results}
\tablehead{
\colhead{Obs ID} &
\colhead{$N_{H}$} &
\colhead{$kT_{bb}$} &
\colhead{$N_{bb}$\tablenotemark{a}} &
\colhead{$\Gamma$} &
\colhead{$N_{\Gamma}$\tablenotemark{b}} &
\colhead{$\chi_{\nu}^{2}/\nu$} \\
\colhead{} &
\colhead{(10$^{22}$ cm$^{-2}$)} &
\colhead{(keV)} &
\colhead{} &
\colhead{} &
\colhead{} &
\colhead{}
}
\startdata
\sidehead{Rising portion of X-ray flare (MJD 55166.4 -- 55177.6)}
00035066002 & 0.3$^{+0.1}_{-0.1}$ & 2.2$^{+0.2}_{-0.3}$ & 0.2$^{+0.1}_{-0.1}$ & 0.9$^{+0.4}_{-0.2}$ & 0.2$^{+0.1}_{-0.1}$ & 1.04/536 \\
00035066005 & 0.3$^{+0.1}_{-0.1}$ & 2.2$^{+0.1}_{-0.1}$ & 0.3$^{+0.1}_{-0.1}$ & 1.2$^{+0.4}_{-0.3}$ & 0.7$^{+0.2}_{-0.1}$ & 1.05/513 \\
00035066006 & 0.3$^{+0.1}_{-0.1}$ & 2.4$^{+0.2}_{-0.3}$ & 0.2$^{+0.1}_{-0.1}$ & 1.0$^{+0.2}_{-0.2}$ & 1.0$^{+0.2}_{-0.2}$ & 1.06/540 \\
00035066007 & 0.8$^{+0.4}_{-0.3}$ & 2.3$^{+0.4}_{-0.7}$ & 0.2$^{+0.2}_{-0.2}$ & 0.8$^{+0.4}_{-0.3}$ & 1.0$^{+0.2}_{-0.2}$ & 1.19/545 \\
00035066008 & 0.3$^{+0.1}_{-0.1}$ & 2.4$^{+0.6}_{-1.3}$ & 0.2$^{+0.3}_{-0.1}$ & 0.6$^{+0.5}_{-0.2}$ & 1.0$^{+0.2}_{-0.2}$ & 1.13/513 \\
00035066009 & 0.3$^{+0.2}_{-0.1}$ & 2.2$^{+0.5}_{-0.4}$ & 0.2$^{+0.2}_{-0.1}$ & 0.6$^{+0.4}_{-0.2}$ & 0.9$^{+0.3}_{-0.2}$ & 1.10/526 \\
\hline \\
\sidehead{Falling portion of X-ray flare (MJD 55178.4 -- 55193.6)}
00035066010 & 0.1$^{+0.2}_{-0.1}$ & 2.0$^{+0.3}_{-0.3}$ & 0.2$^{+0.1}_{-0.1}$ & 0.7$^{+0.6}_{-0.3}$ & 0.6$^{+0.4}_{-0.2}$ & 1.11/330 \\
00035066011 & 0.2$^{+0.4}_{-0.1}$ & 2.5$^{+0.5}_{-0.9}$ & 0.2$^{+0.3}_{-0.1}$ & 0.7$^{+0.8}_{-0.5}$ & 0.9$^{+0.7}_{-0.4}$ & 1.05/488 \\
00035066012 & 0.2$^{+0.1}_{-0.1}$ & 2.2$^{+0.1}_{-0.2}$ & 0.4$^{+0.1}_{-0.1}$ & 1.1$^{+0.3}_{-0.4}$ & 0.8$^{+0.3}_{-0.2}$ & 1.01/517 \\
00035066013 & 0.3$^{+0.2}_{-0.1}$ & 2.2$^{+0.1}_{-0.2}$ & 0.3$^{+0.1}_{-0.1}$ & 1.1$^{+0.4}_{-0.3}$ & 0.9$^{+0.5}_{-0.2}$ & 1.07/550 \\
00035066014 & 0.5$^{+0.2}_{-0.2}$ & 2.4$^{+0.1}_{-0.1}$ & 0.38$^{+0.03}_{-0.06}$ & 1.8$^{+0.6}_{-0.6}$ & 1.4$^{+1.0}_{-0.6}$ & 0.99/491 \\
00035066015 & 0.4$^{+0.2}_{-0.1}$ & 2.0$^{+0.1}_{-0.1}$ & 0.19$^{+0.03}_{-0.04}$ & 1.6$^{+0.5}_{-0.4}$ & 1.0$^{+0.6}_{-0.3}$ & 1.07/506 \\
00035066017 & 0.4$^{+0.1}_{-0.1}$ & 2.1$^{+0.1}_{-0.2}$ & 0.2$^{+0.1}_{-0.1}$ & 1.2$^{+0.5}_{-0.4}$ & 0.9$^{+0.2}_{-0.4}$ & 0.97/528 \\
00035066018 & 0.3$^{+0.1}_{-0.1}$ & 2.1$^{+0.2}_{-0.3}$ & 0.1$^{+0.1}_{-0.1}$ & 0.9$^{+0.4}_{-0.2}$ & 0.7$^{+0.2}_{-0.1}$ & 1.13/547 \\
00035066020 & 0.5$^{+0.3}_{-0.2}$ & 2.2$^{+0.1}_{-0.3}$ & 0.16$^{+0.04}_{-0.07}$ & 1.4$^{+0.7}_{-0.5}$ & 0.8$^{+0.7}_{-0.3}$ & 1.15/361 \\
00035066021 & 0.2$^{+0.2}_{-0.1}$ & 1.8$^{+0.8}_{-0.2}$ & 0.09$^{+0.03}_{-0.02}$ & 0.9$^{+0.4}_{-0.3}$ & 0.3$^{+0.2}_{-0.1}$ & 1.05/369 \\
00035066022 & 0.2$^{+0.1}_{-0.1}$ & 1.7$^{+0.1}_{-0.1}$ & 0.07$^{+0.02}_{-0.01}$ & 1.1$^{+0.5}_{-0.3}$ & 0.2$^{+0.1}_{-0.1}$ & 0.90/407 \\
00035066023 & 1.0$^{+0.1}_{-0.1}$ & 1.7$^{+0.2}_{-0.1}$ & 0.06$^{+0.01}_{-0.01}$ & 0.7$^{+0.2}_{-0.2}$ & 0.20$^{+0.07}_{-0.04}$ & 1.02/489 \\
00035066024 & 0.2$^{+0.3}_{-0.2}$ & 1.7$^{+0.1}_{-0.1}$ & 0.07$^{+0.03}_{-0.01}$ & 0.7$^{+0.4}_{-0.1}$ & 0.3$^{+0.3}_{-0.1}$ & 1.13/333
\enddata
\tablecomments{The columns are: hydrogen column density ($N_H$), blackbody temperature ($T_{bb}$), blackbody normalization ($N_{bb}$), photon index ($\Gamma$), power-law normalization ($N_{\Gamma}$), reduced $\chi_{\nu}^2$ and degrees of freedom $\nu$.}
\tablenotetext{a}{In units of erg s$^{-1}$ kpc$^{-2}$}
\tablenotetext{b}{In units of ph cm$^{-2}$s$^{-1}$keV$^{-1}$}
\end{deluxetable}

\begin{deluxetable}{cccccccc}
\tabletypesize{\scriptsize}
\centering
\tablecolumns{8}
\tablewidth{0pt}
\tablecaption{\label{table:RXTEFits}RXTE/PCA Spectral Results}
\tablehead{
\colhead{Observation ID} &
\colhead{$kT_{bb}$} &
\colhead{$N_{bb}$\tablenotemark{a}} &
\colhead{$kT_{s}$} &
\colhead{$kT_{e}$} &
\colhead{$\tau$} &
\colhead{$N_{comp}$} & 
\colhead{$\chi_{\nu}^{2}/\nu$} \\
\colhead{} &
\colhead{(keV)} &
\colhead{} &
\colhead{(keV)} &
\colhead{(keV)} &
\colhead{} &
\colhead{} &
\colhead{}
}
\startdata
\sidehead{Rising portion of X-ray flare (MJD 55166.4 -- 55177.6)}
94323-04-01-01 & 2.0$_{-0.2}^{+0.3}$ & 0.03$_{-0.01}^{+0.01}$ & 0.4$_{-0.4}^{+0.2}$ & 6.3$_{-0.2}^{+0.3}$ & 8.1$_{-0.5}^{+0.6}$ & 0.4$_{-0.4}^{+0.7}$ & 0.41/40  \\
94323-04-01-00 & 2.3$_{-0.3}^{+0.8}$ & 0.04$_{-0.01}^{+0.02}$ & 0.3$_{-0.3}^{+0.4}$ & 6.5$_{-0.2}^{+0.6}$ & 7.9$_{-0.7}^{+0.6}$ & 0.7$_{-0.2}^{+1.1}$ & 0.23/40 \\
94323-04-01-02 & 2.8$_{-0.6}^{+1.2}$ & 0.07$_{-0.02}^{+0.11}$ & 0.6$_{-0.6}^{+1.1}$ & 6.7$_{-0.4}^{+1.8}$ & 7.7$_{-1.6}^{+0.9}$ & 0.6$_{-0.1}^{+1.4}$ & 0.22/40 \\
94323-04-01-03 & 2.3$_{-0.3}^{+0.6}$ & 0.08$_{-0.01}^{+0.03}$ & 0.3$_{-0.3}^{+0.4}$ & 6.5$_{-0.2}^{+0.5}$ & 8.6$_{-0.7}^{+0.7}$ & 1.0$_{-0.2}^{+1.3}$ & 0.23/40 \\
94323-04-01-04 & 2.3$_{-0.2}^{+0.6}$ & 0.09$_{-0.01}^{+0.02}$ & 0.3$_{-0.3}^{+0.3}$ & 6.5$_{-0.2}^{+0.4}$ & 9.0$_{-0.8}^{+0.8}$ & 1.0$_{-0.1}^{+1.3}$ & 0.28/40 \\
94323-04-01-05 & 2.2$_{-0.2}^{+0.4}$ & 0.09$_{-0.02}^{+0.01}$ & 0.4$_{-0.4}^{+0.3}$ & 6.4$_{-0.2}^{+0.3}$ & 9.3$_{-0.7}^{+0.7}$ & 1.2$_{-0.1}^{+1.3}$ & 0.16/40 \\
94323-04-01-06 & 2.2$_{-0.2}^{+0.4}$ & 0.09$_{-0.02}^{+0.01}$ & 0.4$_{-0.4}^{+0.2}$ & 6.4$_{-0.2}^{+0.3}$ & 9.3$_{-0.7}^{+0.7}$ & 1.1$_{-0.1}^{+1.2}$ & 0.16/40 \\
94323-04-02-00 & 2.3$_{-0.2}^{+0.4}$ & 0.09$_{-0.02}^{+0.02}$ & 0.4$_{-0.4}^{+0.2}$ & 6.5$_{-0.2}^{+0.3}$ & 9.1$_{-0.7}^{+0.7}$ & 1.1$_{-0.1}^{+1.4}$ & 0.25/40 \\
94323-04-02-01 & 2.4$_{-0.3}^{+0.7}$ & 0.10$_{-0.02}^{+0.04}$ & 0.4$_{-0.4}^{+0.2}$ & 6.5$_{-0.2}^{+0.5}$ & 9.0$_{-0.8}^{+0.7}$ & 1.2$_{-0.2}^{+1.5}$ & 0.22/40 \\
\hline \\
\sidehead{Falling portion of X-ray flare (MJD 55178.4 -- 55193.6)}
94323-04-02-02 & 2.3$_{-0.2}^{+0.5}$ & 0.10$_{-0.02}^{+0.02}$ & 0.4$_{-0.4}^{+0.2}$ & 6.5$_{-0.2}^{+0.3}$ & 9.3$_{-0.7}^{+0.7}$ & 1.2$_{-0.1}^{+1.4}$ & 0.25/40 \\
94323-04-02-04 & 2.3$_{-0.2}^{+0.5}$ & 1.10$_{-0.02}^{+0.02}$ & 0.4$_{-0.4}^{+0.2}$ & 6.5$_{-0.2}^{+0.4}$ & 9.2$_{-0.7}^{+0.7}$ & 1.2$_{-0.2}^{+1.4}$ & 0.26/40 \\
94323-05-01-00 & 2.2$_{-0.2}^{+0.4}$ & 0.09$_{-0.01}^{+0.01}$ & 0.4$_{-0.4}^{+0.2}$ & 6.4$_{-0.2}^{+0.3}$ & 9.4$_{-0.7}^{+0.7}$ & 1.1$_{-0.1}^{+1.3}$ & 0.22/40 \\
94323-05-01-01 & 2.2$_{-0.2}^{+0.6}$ & 0.08$_{-0.01}^{+0.01}$ & 0.4$_{-0.4}^{+0.2}$ & 6.4$_{-0.2}^{+0.3}$ & 9.2$_{-0.8}^{+0.6}$ & 1.1$_{-0.1}^{+1.3}$ & 0.24/40 \\
94323-05-01-02 & 2.3$_{-0.3}^{+0.6}$ & 0.08$_{-0.01}^{+0.02}$ & 0.4$_{-0.4}^{+0.3}$ & 6.4$_{-0.2}^{+0.4}$ & 9.1$_{-0.8}^{+0.8}$ & 1.1$_{-0.1}^{+1.3}$ & 0.24/40 \\
94323-05-02-03 & 2.2$_{-0.2}^{+0.4}$ & 0.07$_{-0.01}^{+0.01}$ & 0.4$_{-0.4}^{+0.2}$ & 6.4$_{-0.2}^{+0.3}$ & 9.2$_{-0.7}^{+0.7}$ & 0.9$_{-0.1}^{+1.2}$ & 0.22/40 \\
94323-05-02-00 & 2.2$_{-0.2}^{+0.4}$ & 0.07$_{-0.01}^{+0.10}$ & 0.4$_{-0.4}^{+0.3}$ & 6.4$_{-0.2}^{+0.3}$ & 9.0$_{-0.6}^{+0.7}$ & 0.9$_{-0.1}^{+1.1}$ & 0.15/40 \\
94323-05-02-06 & 2.2$_{-0.2}^{+0.6}$ & 0.06$_{-0.01}^{+0.01}$ & 0.4$_{-0.4}^{+0.3}$ & 6.4$_{-0.2}^{+0.3}$ & 8.9$_{-0.6}^{+0.7}$ & 0.9$_{-0.1}^{+1.2}$ & 0.20/40 \\
94323-05-02-04 & 2.1$_{-0.2}^{+0.5}$ & 0.05$_{-0.01}^{+0.01}$ & 0.3$_{-0.3}^{+0.4}$ & 6.4$_{-0.2}^{+0.3}$ & 8.9$_{-0.7}^{+0.7}$ & 0.9$_{-0.1}^{+1.1}$ & 0.17/40 \\
94323-05-02-01 & 2.1$_{-0.2}^{+0.4}$ & 0.05$_{-0.01}^{+0.01}$ & 0.4$_{-0.4}^{+0.3}$ & 6.4$_{-0.2}^{+0.3}$ & 8.9$_{-0.5}^{+0.6}$ & 0.8$_{-0.1}^{+1.0}$ & 0.14/40 \\
94323-05-02-02 & 2.1$_{-0.2}^{+0.5}$ & 0.04$_{-0.07}^{+0.08}$ & 0.3$_{-0.3}^{+0.3}$ & 6.4$_{-0.2}^{+0.3}$ & 8.4$_{-0.5}^{+0.6}$ & 0.7$_{-0.1}^{+1.0}$ & 0.15/40 \\
94323-05-02-05 & 2.1$_{-0.2}^{+0.4}$ & 0.03$_{-0.01}^{+0.01}$ & 0.4$_{-0.4}^{+0.3}$ & 6.4$_{-0.2}^{+0.3}$ & 8.4$_{-0.5}^{+0.6}$ & 0.6$_{-0.1}^{+0.9}$ & 0.17/40 \\
94323-05-03-00 & 2.1$_{-0.2}^{+0.5}$ & 0.03$_{-0.01}^{+0.01}$ & 0.3$_{-0.3}^{+0.4}$ & 6.4$_{-0.2}^{+0.3}$ & 8.4$_{-0.6}^{+0.7}$ & 0.6$_{-0.1}^{+0.9}$ & 0.30/40 \\
94323-05-03-01 & 2.0$_{-0.1}^{+0.3}$ & 0.03$_{-0.01}^{+0.01}$ & 0.4$_{-0.4}^{+0.2}$ & 6.2$_{-0.1}^{+0.2}$ & 8.5$_{-0.5}^{+0.5}$ & 0.44$_{-0.04}^{+0.73}$ & 0.20/40 \\
94323-05-03-02 & 2.0$_{-0.2}^{+0.4}$ & 0.01$_{-0.01}^{+0.01}$ & 0.4$_{-0.4}^{+0.3}$ & 6.2$_{-0.2}^{+0.3}$ & 8.3$_{-0.5}^{+0.7}$ & 0.4$_{-0.1}^{+0.6}$ & 0.20/40 \\
94323-05-03-03 & 2.0$_{-0.2}^{+0.4}$ & 0.02$_{-0.01}^{+0.01}$ & 0.3$_{-0.3}^{+0.4}$ & 6.1$_{-0.2}^{+0.3}$ & 8.2$_{-0.6}^{+0.7}$ & 0.4$_{-0.1}^{+0.6}$ & 0.20/40 \\
\hline \\
\sidehead{Apastron (MJD 55199.4 -- 55216.6)}
94323-05-04-00 & 1.9$_{-0.2}^{+0.4}$ & 0.007$_{-0.002}^{+0.002}$ & 0.4$_{-0.4}^{+0.2}$ & 6.3$_{-0.2}^{+0.4}$ & 6.9$_{-0.5}^{+0.5}$ & 0.15$_{-0.02}^{+0.43}$ & 0.61/40 \\
94323-05-04-01 & 1.8$_{-0.2}^{+0.7}$ & 0.004$_{-0.001}^{+0.002}$ & 0.4$_{-0.4}^{+0.3}$ & 6.4$_{-0.4}^{+0.7}$ & 6.3$_{-0.7}^{+0.6}$ & 0.09$_{-0.02}^{+0.13}$ & 0.61/40 \\
94323-05-04-02 & 2.6$_{-1.0}^{+1.2}$ & 0.002$_{-0.001}^{+0.001}$ & 0.6$_{-0.6}^{+0.1}$ & 3.4$_{-1.2}^{+2.3}$ & 3.6$_{-2.5}^{+1.7}$ & 0.2$_{-0.2}^{+0.1}$ & 0.67/40 \\
\enddata
\tablecomments{The parameters for the Compton component are: seed photon temperature ($T_s$), electron temperature ($T_e$), optical depth ($\tau$), and normalization ($N_{comp}$). The geometry parameter is frozen at 0.8, and $N_{H}$ is fixed at $0.3\times10^{22}$cm$^{-2}$. }
\tablenotetext{a}{In units of erg s$^{-1}$ kpc$^{-2}$}
\end{deluxetable}

\begin{sidewaystable}[h]
\renewcommand{\arraystretch}{1.0} \renewcommand{\tabcolsep}{0.2cm}\fontsize{10.0}{14.5}\selectfont
{\centering
\caption{\label{table:JointSwift/RXTE}Joint Swift/RXTE Spectral Fits}
\begin{tabular}{ccccccccccc} 
\hline\hline
& $N_{H} $ & $kT_{bb}$ & $N_{bb}$$^{\dag}$ & $kT_{bb}$ & $N_{bb}$$^{\dag}$ & $kT_{s}$ & $kT_{e}$ & $\tau$ & $N_{comp}$ & $\chi_{\nu}^{2}/\nu$ \\
& $\:10^{22}\mathrm{cm}^{-2}$ & (keV) &  & (keV) &  & (keV) & (keV) & & & \\
\cline{3-6} & & \multicolumn{2}{c}{Swift} & \multicolumn{2}{c}{RXTE} & & & & &  \\
\hline
\multirow{2}{*}{\textit{Rising portion}} & 0.34$_{-0.02}^{+0.01}$ & 1.95$_{-0.07}^{+0.06}$ & 0.27$_{-0.03}^{+0.01}$ & 2.02$_{-0.12}^{+0.16}$ & 0.08$^{+0.02}_{-0.02}$ & 0.11$_{-0.11}^{+0.05}$ & 6.3$^{+0.2}_{-0.1}$ & 9.5$^{+0.6}_{-0.5}$ & 1.20$^{+0.83}_{-0.10}$ & 1.18/851 \\
\cline{3-6} & 0.33$^{+0.01}_{-0.01}$ & \multicolumn{4}{c}{1.65$^{+0.03}_{-0.03}{}^{\ddagger}\quad0.168^{+0.007}_{-0.006}$$^{\ddagger}$} & 0.132$^{+0.04}_{-0.13}$ & 5.87$^{+0.07}_{-0.07}$ & 12.31$^{+0.2}_{-0.6}$ & 1.6$^{+0.7}_{-0.1}$ & 1.44/853 \\
\hline
\multirow{2}{*}{\textit{Falling portion}} & 0.33$^{+0.01}_{-0.01}$ & 1.94$_{-0.04}^{+0.05}$ & 0.118$^{+0.006}_{-0.006}$ & 2.33$^{+0.58}_{-0.26}$ & 0.017$^{+0.203}_{-0.004}$ & 0.06$^{+0.18}_{-0.06}$ & 6.5$^{+0.3}_{-0.2}$ & 8.0$^{+0.4}_{-0.5}$  & 0.5$^{+0.3}_{-0.1}$ & 1.32/881 \\
\cline{3-6}  & 0.24$^{+0.03}_{-0.03}$ & \multicolumn{4}{c}{1.49$^{+0.02}_{-0.02}{}^{\ddagger}\quad0.071^{+0.002}_{-0.002}$$^{\ddagger}$} & 0.23$^{+0.04}_{-0.04}$ & 5.67$^{+0.05}_{-0.05}$ & 12.70$^{+0.3}_{-0.4}$ &  0.52$^{+0.02}_{-0.02}$ & 1.78/877 \\
\hline
\hline \\
\end{tabular} 
}
\fontsize{8.0}{8.5}\selectfont{Note. --- The blackbody component is shown for Swift/XRT and RXTE/PCA data separately. The parameters for the Compton component are: seed photon temperature ($T_s$), electron temperature ($T_e$), optical depth ($\tau$), and normalization ($N_{comp}$). The geometry parameter is frozen at 0.8.\\
$^{\dag}${In units of erg s$^{-1}$ kpc$^{-2}$} \\
$^{\ddagger}${The blackbody temperature $kT_{bb}$ and normalization parameter $N_{bb}$, respectively, are tied between the Swift and RXTE data.}}

\label{table:Swift-RXTE}
\end{sidewaystable}

\clearpage

\end{document}